%% file: arxiv.tex
\documentclass{article} 
\usepackage{iclr2026_conference_arxiv,times}

\input{math_commands.tex}

\usepackage{hyperref}
\usepackage{url}
\usepackage{graphicx} 
\usepackage{booktabs} 
\usepackage{multirow} 
\usepackage{graphicx}
\usepackage{subcaption}
\usepackage{wrapfig}

\usepackage{color,soul}

\usepackage{xcolor}

\title{AgentSoV: Virtual Sensing for LLM Multi-Agent Systems via Stream of Variation}

\author{Congjing Zhang$^{1}$ \quad
Siyuan Xu$^{2}$ \quad
Yunfei Bai$^{3}$ \quad 
Ke Ren$^{3}$ \quad 
Xiaoning Qian$^{2}$ \quad \\[8pt]
\textbf{Shuai Huang$^{1}$}\thanks{Corresponding author.} \\[8pt]
$^{1}$Department of Industrial \& Systems Engineering, University of Washington \\
$^{2}$Department of Electrical \& Computer Engineering, Texas A\&M University \\
$^{3}$Amazon \\[6pt]
\texttt{congjing@uw.edu \quad siyuanxu@tamu.edu \quad byunfei@amazon.com}\\
\texttt{renkea@amazon.com \quad xqian@ece.tamu.edu \quad shuaih@uw.edu}
}

\iclrarxivcopy
\renewcommand{\headrulewidth}{0pt}
\begin{document}

\maketitle

\begin{abstract}

Large language model (LLM) multi-agent systems (MASs) solve complex tasks through interacting agents, yet their reliability is difficult to analyze because failures depend on both individual agents and how information and errors propagate across the system. Existing approaches largely treat this as post-hoc failure attribution, but do not provide a common model across heterogeneous MAS architectures or support broader analysis of information influence and observability. We introduce \textbf{AgentSoV}, a virtual-sensing framework that models MAS execution as a stochastic dynamic system through Stream of Variation (SoV). AgentSoV maps heterogeneous trajectories into diagnostic stages, constructs virtual measurements from observable outputs, and learns cross-stage variation propagation with a state-space model. Its representation is architecture-aware in stage dependencies yet architecture-agnostic in analytical language, supporting chain-structured, orchestration-based, and directed acyclic graph (DAG)-structured MASs. Without re-executing the MAS, AgentSoV supports three system-level analyses: \emph{failure attribution} identifies where abnormal variation enters; \emph{stage-input influence} quantifies how strongly intermediate inputs affect downstream outcomes; and \emph{virtual-sensor placement} selects complementary observations under a monitoring budget. Experiments across multiple MAS architectures demonstrate a unified approach to fault localization, information influence, and observability. AgentSoV reframes agentic execution as a dynamic system to model rather than merely text to inspect, providing insights for MAS structure and component design.

\end{abstract}

\section{Introduction}
\label{sec:intro}
Large language model (LLM) multi-agent systems (MASs) solve complex tasks by distributing work across specialized agents \citep{li2024survey,yang2026agentnet}. These agents may plan, call tools, or verify one another's outputs. However, composing capable agents does not necessarily yield a reliable system. Failures can arise from individual agent errors, but also from system design and ineffective information exchange \citep{cemri2026multi}. 
Intermediate outputs may have little effect on the final result or duplicate information already available elsewhere. They may also introduce erroneous variation that persists through subsequent interactions \citep{zhang2026unlocking,li2025agent}. MAS reliability is therefore a system-level problem. Effective diagnosis should locate where a failure originates and trace how its effect propagates. It should also reveal which intermediate information matters to the outcome and which parts of the execution need to be observed.


A major line of work studies this problem through failure attribution:
given a completed failed trajectory, identify the agent or execution step
responsible for the failure \citep{zhang2026agentracer}. Who\&When \citep{pmlr-v267-zhang25cq} formalized agent- and step-level attribution over
failure trajectories from a large collection of MASs and let LLM judges inspect trajectories.
Subsequent approaches improve particular parts of this pipeline by narrowing
the scope before localization, training specialized failure tracers, refining
trajectory inspection, or replacing direct judging with intervention,
counterfactual replay, and spectrum analysis
\citep{ma2026dover,yeh2026tracing,li2026beyond}.
For example, on the Who\&When benchmark, SDBL \citep{sun2026scope} reports step-level accuracy up
to $27.78\%$ on the handcrafted subset and $33.33\%$ on the automatically
generated subset, and AgenTracer \citep{zhang2026agentracer} reports up to $20.68\%$ and $42.86\%$,
respectively.
These results represent substantial progress, but they also illustrate that
fine-grained MAS diagnosis remains difficult. Additional related work is in Appendix \ref{app_rel_work}.

More fundamentally, many existing approaches are not architecture-invariant by construction. The meaning of an execution step depends on how the MAS is implemented. In an orchestration-based system, a step may represent an orchestrator-worker interaction, while in a  directed acyclic graph (DAG)-structured workflow it may correspond to a node with multiple upstream parents.
Methods that reason directly over these execution objects are therefore tied to the underlying trajectory format. 
Intervention-based methods may also require changes to the replay mechanism, while learned approaches may depend on a new training distribution \citep{zhang2026agentracer}. Replay-based methods further require access to a re-executable MAS and its environment \citep{yeh2026tracing}. Although such methods can be adapted across systems, their diagnostic representation remains closely coupled to the execution format.
Most of these methods are also designed primarily for failure attribution. They answer where or by whom a failure was introduced, but do not provide a common model for broader system analysis. For example, they do not directly characterize how strongly variation persists downstream. 
They also do not determine which intermediate inputs or outputs are redundant, nor which should be monitored for effective diagnosis.


These questions suggest a different starting point: an MAS execution can be
viewed as a stochastic dynamic system. This viewpoint has a close
analogue in multistage manufacturing. Manufacturing systems are highly
heterogeneous, i.e., different processes contain different machines, operations, and sensor layouts, yet state-space models provide a
common abstraction in which a latent process state evolves across stages,
process inputs affect that state, and measurements reveal only part of it.
One representative framework is Stream of Variation (SoV), which models how
variation introduced at one manufacturing stage propagates through downstream
stages and appears in distributed measurements
\citep{shi2006stream,huang2000stream,zhou2003diagnosability}.
Once the process is represented in this form, concepts from control and
systems theory become available: observability asks whether internal
variation can be inferred from available measurements,
diagnosability asks whether different variation sources can be
distinguished, and sensor-design methods ask where measurements should be
placed to make the process sufficiently informative for diagnosis.
The value of this abstraction is not that every manufacturing process has the
same dynamics, but that very different processes can be represented using the
same language of \emph{stages, propagated state, inputs, disturbances, and
measurements}. We discuss a more detailed illustration in Section \ref{sec:illu}.

\begin{figure}[t]
\begin{center}
\includegraphics[width=1\linewidth]{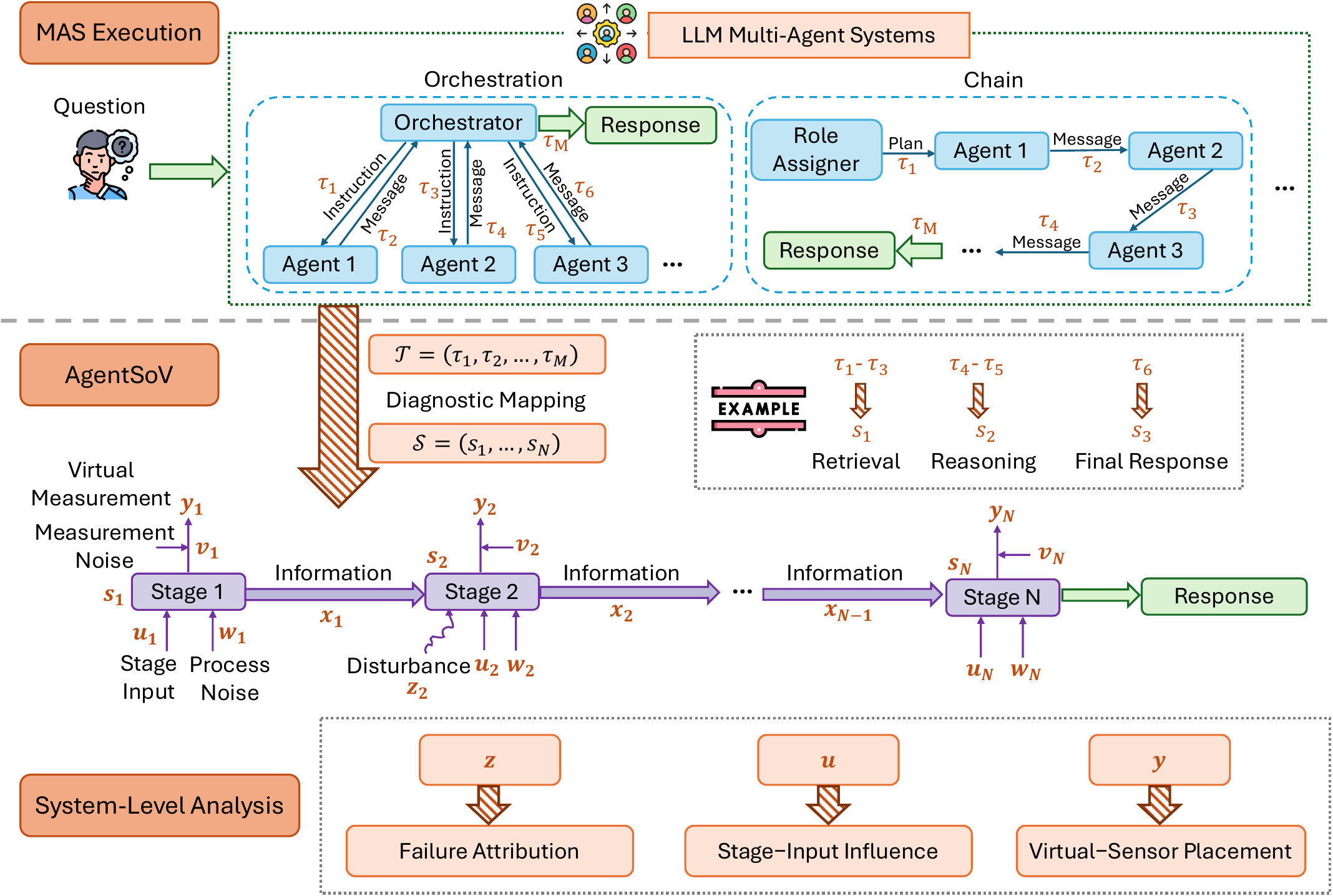}
\end{center}
\caption{Schematic overview of AgentSoV. The upper panel shows the observed LLM MAS execution trajectory $\mathcal{T}$. AgentSoV maps $\mathcal{T}$ into diagnostic stages $\mathcal{S}$, constructs stage inputs $\vu$ and virtual measurements $\vy$, and models cross-stage variation through SoV dynamics. The diagnostic mapping need not preserve the original execution one-to-one: multiple execution steps $\tau$ may be grouped into a stage $s$ according to their diagnostic role. The resulting representation supports three system-level analyses through different model quantities: disturbances $\vz$ for failure attribution, stage inputs $\vu$ for stage-input influence, and virtual measurements $\vy$ for virtual-sensor placement.}
\label{fig_exe_diag}
\end{figure}

Building on this connection, we introduce \textbf{AgentSoV}, a virtual-sensing
framework that models LLM MAS execution through SoV
(Figure~\ref{fig_exe_diag}). AgentSoV maps an observed execution into a
sequence of diagnostic stages, constructs process inputs and virtual
measurements from observable execution information, and fits state-space
dynamics that characterize cross-stage variation propagation. Once learned,
the same compact model supports failure attribution, stage-input influence
analysis, and virtual-sensor placement without requiring a separate diagnostic
representation for each task. We instantiate this framework for
chain-structured, orchestration-based, and DAG-structured MASs, demonstrating
that the diagnostic mapping can change with the execution architecture while
the underlying analytical framework remains the same. In summary, our contributions are:
(1) we introduce AgentSoV, a virtual-sensing framework that represents LLM MAS
execution as a multistage dynamic system using SoV, providing a common
statistical language for modeling information and variation propagation;
(2) we show that the resulting dynamics support multiple system-level analyses
within a single model, including failure attribution, stage-input influence, and virtual-sensor placement;
(3) we adapt AgentSoV to chain-structured, orchestration-based, and
DAG-structured MASs, demonstrating that its diagnostic stages and dependencies
can accommodate heterogeneous execution architectures without requiring a
one-to-one reproduction of the underlying trajectory; and
(4) we validate AgentSoV across these architectures, demonstrating effective
diagnosis from a unified learned representation without
re-executing the underlying MAS.

\section{AgentSoV Framework}
\label{sec:framework}
\subsection{Illustration of the State-Space Representation for MASs} \label{sec:illu}
Figure \ref{fig_exe_diag} shows an overview of our AgentSoV framework. Consider an MAS in which an Orchestrator asks a WebSurfer to retrieve
evidence, passes the result to a Coder for analysis, and then produces a final
response. Instead of treating every message as a separate diagnostic object,
we represent this execution by three functional stages: $s_1$ (retrieval),
$s_2$ (reasoning), and $s_3$ (final response). At stage $k, k\in \{1, 2, 3\}$, the vector
$\vu_k$ contains the information supplied to the stage, $\vx_k$ is a latent
diagnostic state that carries task-relevant variation, $\vz_k$ is an optional
disturbance vector representing abnormal variation introduced at the stage, and $\vy_k$ is a virtual measurement constructed from the observable stage output, such as a text representation or a tool-execution statistic. Conceptually,
\begin{align}
    \vx_k
    &= \mF_k\vx_{k-1}+\mB_k\vu_k+\mH_k\vz_k+\vw_k\qquad
 \text{(State equation)},
    \label{eq:state_equation_s3}\\
    \vy_k&=\mC_k\vx_k+\vv_k\qquad
 \text{(Observation equation)}.
    \label{eq:observation_equation_s3}
\end{align}
Here, the matrices determine how the
configured variables are related: $\mF_k$ propagates upstream variation, $\mB_k$ maps the current input,
$\mH_k$ maps the abnormal disturbance into the diagnostic state $\vx_k$, and $\mC_k$ specifies which components of the diagnostic state are visible through the virtual measurement $\vy_k$. If the
WebSurfer retrieves an incorrect fact, for example, that abnormal variation is
represented by the disturbance $\mH_k \vz_k$ and may subsequently propagate
through the reasoning stage $s_2$ into the final answer. The process noise $\vw_k$
captures unmodeled stage-to-stage variation, whereas the measurement noise
$\vv_k$ captures uncertainty in constructing $\vy_k$ from an observable
output (e.g., an imperfect diagnostic feature). Thus, an agent response acts like a location in a manufacturing
process at which a virtual sensor can be constructed
\citep{kabadayi2006virtual}. Stage-indexed matrices allow heterogeneous stage
operations; when stages share compatible dynamics, AgentSoV instead uses
shared matrices such as $\mF$, $\mB$, and $\mC$.

The stages form an analytical abstraction of information transformation and
need not reproduce the execution log one-to-one. A stage may combine an
assistant turn with its tool calls and responses, several raw operations may
be aggregated, and irrelevant operations may be omitted. The dependency
structure follows the MAS architecture: sequential stages depend on their
predecessors, an Orchestrator's planning state may affect several workers, and
a node in a DAG may depend on multiple parents. AgentSoV is
therefore architecture-aware when constructing the diagnostic representation
while retaining a common analytical language across architectures.

The same dynamics support three system-level analyses. \emph{Failure
attribution} identifies the disturbance source $\vz_k$ whose propagated signature best
explains the observed residuals. \emph{Stage-input influence} changes
$\vu_k$ and propagates the resulting change in $\mB_k\vu_k$ through the fitted
dynamics, revealing whether information supplied to an intermediate stage
affects a later output
\citep{li2025agent,zhang2026unlocking}. \emph{Virtual-sensor placement}
selects a compact set of intermediate measurements $\vy_k$ that makes
candidate disturbance sources $\vz_k$ observable and distinguishable when monitoring
every output is unnecessary or costly.
This correspondence defines the roles of $\vu_k$, $\vx_k$, $\vy_k$, and
$\vz_k$. We next describe how execution operations are mapped into diagnostic
stages, after which we specify and estimate the resulting AgentSoV model.

\subsection{From Execution Trajectories to Diagnostic Stages}
\label{sec:trace_to_stages}

Let an execution trajectory consist of $M$ ordered steps
$\mathcal{T}=(\tau_1,\ldots,\tau_M)$, such as agent messages, tool calls, or planning operations. AgentSoV maps
these operations into $N$ diagnostic stages:
\begin{equation}
\mathcal{T}\rightarrow
\mathcal{S}=(s_1,\ldots,s_N),
\label{eq:stages}
\end{equation}
where each $s_k$, $k\in\{1,\ldots,N\}$, groups execution operations that
play a common diagnostic role. The mapping need not preserve the raw
execution one-to-one: several operations may be aggregated into one stage,
while operations irrelevant to the diagnostic objective may be omitted.
Consider again the Orchestrator-WebSurfer-Coder example from
Section~\ref{sec:illu}, answering the question ``Which two ASEAN countries
have capitals that are furthest apart?'' A simplified trajectory contains
$\tau_1$: the Orchestrator constructs a fact sheet and plan;
$\tau_2$: the Orchestrator instructs the WebSurfer to retrieve the relevant
capitals;
$\tau_3$: the WebSurfer returns a candidate pair;
$\tau_4$: the Orchestrator passes the retrieved information to the Coder;
$\tau_5$: the Coder constructs the answer; and
$\tau_6$: the Orchestrator delivers the final response.
Rather than treating all six operations as separate diagnostic objects,
AgentSoV may represent the execution as
\begin{equation}
\underbrace{(\tau_1,\tau_2,\tau_3)}_{s_1:\,\text{retrieval}}
\rightarrow
\underbrace{(\tau_4,\tau_5)}_{s_2:\,\text{reasoning}}
\rightarrow
\underbrace{\tau_6}_{s_3:\,\text{final response}}.
\label{eq:example_stages}
\end{equation}
The important modeling decision is therefore not simply which agent produces
each raw message, but which execution operations form a meaningful unit of
information transformation and variation propagation. Once these stages are
chosen, the quantities introduced in Section~\ref{sec:illu} (e.g., $\vu_k$, $\vx_k$,
$\vy_k$), and, when explicitly modeled, $\vz_k$ are instantiated from the
information entering, carried through, and observed at each stage. Their exact
construction can vary with the execution architecture without changing their
roles in the AgentSoV representation. More modeling examples are given in Appendix \ref{sec:adaptation} for different MAS architectures.

\subsection{Model Configuration}
\label{sec:model_configuration}

Eqs.~\ref{eq:state_equation_s3}-\ref{eq:observation_equation_s3}
define the state-space representation in AgentSoV. Model configuration assigns meanings
and dimensions to its variables and determines which parameters are fixed,
shared, or learned. We assume isotropic Gaussian noise,
$\vw_k\sim\mathcal{N}(\mathbf{0},\sigma_{w,k}^2\mI)$ and
$\vv_k\sim\mathcal{N}(\mathbf{0},\sigma_{v,k}^2\mI)$, where
$\sigma_{w,k}^2$ and $\sigma_{v,k}^2$ are the stage-specific process- and
measurement-noise variances, respectively. AgentSoV learns the transition
matrices $\mF_k$, input matrices $\mB_k$, $\sigma_{w,k}^2$, and $\sigma_{v,k}^2$ from training
trajectories. The measurement matrix $\mC_k$ is fixed when determined by the
virtual-measurement design and may otherwise be estimated. The disturbance
matrix $\mH_k$ is specified from system knowledge or estimated from controlled
perturbations or labeled failures when structured disturbances are explicitly
modeled. Otherwise, we set $\vz_k=\mathbf{0}$ and fit the normal dynamics.
Although we use stage-indexed matrices for generality, common matrices such as
$\mF$, $\mB$, and $\mC$ can be used when stages perform comparable operations
and the available data support parameter sharing.

Returning to the running example makes the distinction among these terms
concrete. Suppose $\vx_k =
\begin{bmatrix}
x_k^{\mathrm{ret}} &
x_k^{\mathrm{reas}} &
x_k^{\mathrm{ver}}
\end{bmatrix}^{\!\top}$
summarizes retrieval, reasoning, and verification variation. At the retrieval
stage, the task and Orchestrator instruction form the information represented
by $\vu_1$, while the WebSurfer response provides the observable signal from
which $\vy_1$ is constructed. If the WebSurfer incorrectly returns
Jakarta-Bangkok, the resulting abnormal retrieval variation can be represented
through the disturbance channel $\mH_1\vz_1$. For illustration, let
\begin{equation}
\vz_k =
\begin{bmatrix}
z_k^{\mathrm{ret}} \\
z_k^{\mathrm{reas}}
\end{bmatrix},
\qquad
\mH_k =
\begin{bmatrix}
1   & 0 \\
0.3 & 1 \\
0.2 & 0
\end{bmatrix},
\quad
\mF_2 =
\begin{bmatrix}
0.9 & 0   & 0\\
0.3 & 0.5 & 0\\
0.2 & 0   & 0.6
\end{bmatrix}.
\label{eq_mat_example}
\end{equation}
Under this specified disturbance basis, the two coordinates of $\vz_k$ define
retrieval and reasoning disturbances, and the corresponding columns of
$\mH_k$ specify the directions in diagnostic-state space along which those
disturbances enter.
At $s_2$, the Coder receives its current instruction and retrieved evidence
through $\vu_2$. Its state depends on this current-stage information through
$\mB_2\vu_2$ and on variation inherited from the retrieval stage through
$\mF_2\vx_1$. The matrix $\mF_2$ in (\ref{eq_mat_example}) describes
strong persistence of retrieval variation while allowing it to propagate into
the reasoning and verification components. A downstream stage that
independently checks and corrects the retrieved information would instead have
transition dynamics that attenuate this variation. Thus, $\mF_k$ captures
whether upstream variation is preserved, amplified, or suppressed, whereas
$\mB_k$ captures how current-stage information affects the diagnostic state.
This separation is later used to distinguish variation propagation from
stage-input influence.
Finally, $\mC_k$ connects the diagnostic state $\vx_k$ to the retained virtual
measurement $\vy_k$. When the measurement directly represents the state,
$\mC_k=\mI$; when only selected features of an agent response are retained,
$\mC_k$ selects or combines the corresponding state components. This
distinction is especially important for the virtual-sensor analysis in
Section~\ref{sec:sensor_placement}, where different observable stage outputs
provide different amounts of diagnostic information.

\subsection{Model Estimation}
\label{sec:model_estimation}

Once the diagnostic stages $\mathcal{S}$, variable representations, dependency
structure, and parameter configuration have been specified, estimating
AgentSoV is a standard state-space identification problem. For latent-state
instantiations, we fit the parameters designated as learnable using
expectation-maximization (EM). In the E-step, Kalman filtering and
Rauch-Tung-Striebel smoothing infer the latent states given the current
parameter estimates. In the M-step, these parameters are updated from the
inferred state transitions and measurement residuals.
When the diagnostic state $\vx_k$ is directly observable, for example when
$\vx_k=\vy_k$ is a projected representation of an agent response, estimation
reduces to fitting the corresponding cross-stage regression. We use normal,
i.e., fault-free, training executions to characterize the normal dynamics,
corresponding to $\vz_k=\mathbf{0}$. 

Thus, the contribution of AgentSoV is not
a new estimator for linear state-space models. Rather, it is the construction
of a diagnostic representation of MAS execution for which established
state-space estimation and systems-analysis tools become applicable.

\subsection{Instantiating AgentSoV across MAS Architectures}
Our formulation in AgentSoV is not tied to a particular MAS architecture. What changes across systems is how the diagnostic stages and dependencies are instantiated, not the roles of the AgentSoV variables themselves. In a chain-structured system, variation propagates primarily from the preceding stage; in an orchestration-based system, a persistent planning state may influence multiple downstream stages; and in a DAG-structured system, a stage may aggregate variation from several upstream parents. We provide concrete AgentSoV instantiations for these three settings, including BFCL \citep{patil2025berkeley}, Magentic-One \citep{pmlr-v267-zhang25cq}, and AEGIS \citep{kong2026aegis}, in Appendix~\ref{sec:adaptation}. These examples show that the trajectory-to-stage mapping and transition structure can adapt to the underlying execution architecture while retaining the same analytical language of inputs $\vu$, diagnostic states $\vx$, virtual measurements $\vy$, disturbances $\vz$, and variation propagation.

\section{System-Level Analysis with AgentSoV}
\label{sec:diagnosis}

Once the AgentSoV dynamics in
Eqs.~\ref{eq:state_equation_s3}-\ref{eq:observation_equation_s3}
have been fitted, the model can be used not only to predict how an MAS normally evolves, but also to reason about how variation enters, propagates,
and becomes observable. 
We consider three complementary questions: where does abnormal variation originate, how strongly does input supplied at a stage affect downstream outcomes, and which intermediate outputs must be observed for effective diagnosis? All three analyses use
the same fitted state-space dynamics and therefore do not require re-executing
the underlying MAS.

\subsection{Failure Attribution}
\label{sec:fault_attribution}
Failure attribution asks which stage introduced the abnormal variation. Under normal execution, we set
$\vz_k=\mathbf{0}$ in Eq.~\ref{eq:state_equation_s3} and use the fitted model
to predict the stage outputs. A disturbance $\vz_k$ introduced at stage $k$
enters the diagnostic state through $\mH_k\vz_k$. Recursive application of
Eq.~\ref{eq:state_equation_s3} shows that its effect on the virtual
measurement $\vy_l$ at a downstream stage $\ell>k$ is
\begin{equation}
\underbrace{
\mC_\ell\mF_\ell\mF_{\ell-1}\cdots\mF_{k+1}\mH_k
}_{\mG_{\ell,k}}
\vz_k.
\label{eq:disturbance_signature}
\end{equation}
$\mG_{\ell,k}=\mC_\ell\mF_\ell\mF_{\ell-1}\cdots\mF_{k+1}\mH_k$ is
the propagated disturbance signature from stage $k$ to stage $\ell$, a term established in state-space model-based diagnosis
\citep{gertler1998fault,isermann2006fault}. It combines three elements of the AgentSoV model:
$\mH_k$ specifies where abnormal variation enters,
the transition matrices $\mF_{k+1},\ldots,\mF_\ell$ describe how it propagates,
and $\mC_\ell$ specifies how that propagated variation becomes observable.
Accordingly, $\mG_{\ell,k}\vz_k$ is the observable effect of disturbance
$\vz_k$ at stage $\ell$.

In the Orchestrator-WebSurfer-Coder example, an incorrect retrieval
introduced at $s_1$ should influence both the Coder response at $s_2$ and,
unless corrected, the final response at $s_3$. A reasoning error introduced
only at $s_2$ cannot explain abnormal behavior already visible at $s_1$. AgentSoV therefore attributes failure by comparing the observed deviations from normal predictions with the downstream patterns implied by candidate
$\mH_k\vz_k$ terms. When several stages may contribute, the candidate
disturbances can be estimated jointly with a sparsity penalty for accurate identification (see Eq. \ref{eq:app_sparse_attribution} in Appendix~\ref{app:sparse_attribution}). This differs from local anomaly scoring since an attributed source must explain
how its variation propagates through the execution, not merely exhibit a
large local error. The discussion describes the attribution principle at the system level. For completeness, Appendix~\ref{app:residual_diagnosis} gives the corresponding computational formulation.

\subsection{Stage-Input Influence}
\label{sec:information_requirement}

Failure attribution concerns the disturbance channel $\mH_k\vz_k$ in
Eq.~\ref{eq:state_equation_s3}. The process-input channel $\mB_k\vu_k$
supports a different question: how strongly does the information
entering at stage $k$ affect a later measurement? This question applies to
successful as well as failed trajectories. For example, the WebSurfer may
return correct information that the Coder largely ignores, or information
that is redundant with evidence available elsewhere
\citep{li2025agent,zhang2026unlocking}. To quantify this effect, we set the observed input $\vu_k$ to zero in the
fitted model while holding the other stage inputs fixed. Let
$\hat{\vy}_\ell^{-k}$ denote the predicted measurement at a later stage
$\ell$ under this intervention. The linear state and observation equations,
Eqs.~\ref{eq:state_equation_s3}-\ref{eq:observation_equation_s3}, give
\begin{equation}
\hat{\vy}_\ell-\hat{\vy}_\ell^{-k}
=\mC_\ell\mF_\ell\mF_{\ell-1}\cdots\mF_{k+1}\mB_k\vu_k,
\qquad
\quad k<\ell. \label{eq:stage_to_stage_sensitivity}
\end{equation}
$\|\hat{\vy}_\ell-\hat{\vy}_\ell^{-k}\|_2$ is the influence magnitude of the predicted change at stage $\ell$
caused by removing the modeled input at stage $k$. Taking $\ell=N$ measures
its effect on the final measurement. A larger influence $\|\hat{\vy}_\ell-\hat{\vy}_\ell^{-k}\|_2$ means that the input is more consequential. No MAS replay is
required.

\subsection{Virtual-Sensor Placement}
\label{sec:sensor_placement}

The preceding analyses assume that intermediate outputs $\vy_k$ are available
for diagnosis. In practice, monitoring every stage may be unnecessary:
different outputs can reveal nearly the same propagated variation, while
omitting another output may make important failure sources difficult to
distinguish. AgentSoV therefore treats the choice of which $\vy_k$ to retain as
a virtual-sensor placement problem.
The propagated disturbance signature $\mG_{\ell,k}$ gives the key quantity. Observing
$\vy_\ell$ is useful for diagnosing a disturbance at stage $k$ when
$\mG_{\ell,k}\vz_k$ is large relative to normal variation.
More importantly, two candidate measurements are redundant when they expose
essentially the same disturbance directions. AgentSoV therefore selects a
limited set of observation stages $\mathcal S$ using the standard log-determinant information criterion, $\mathcal S^\star=\arg\max_{|\mathcal S|\le b}\log\det\!\left(\sum_{\ell\in\mathcal S}
\mathcal I_\ell+\eta\mI\right),$ where $\mathcal I_\ell$ is the information about the candidate propagated disturbances contained in $\vy_\ell$ relative to normal residual
variation, $b$ is the monitoring budget, and $\eta>0$ stabilizes the
calculation. $\mathcal I_\ell$ is computed from
the propagated effects in $\mG_{\ell,k}$ and the residual covariance estimated from fault-free trajectories.

Returning to the toy example, observing only the final response may not distinguish an incorrect WebSurfer retrieval from an error introduced by the Coder, because both may produce the same final answer. Adding the WebSurfer
output can expose the retrieval error before reasoning occurs, whereas adding
another downstream output $\vy$ carrying nearly identical information may provide
little additional diagnostic value. Virtual-sensor placement therefore asks
not simply where can we observe the MAS, but which observations make its sources of variation most distinguishable?

\vspace{-3pt}

\section{Experiments}

\vspace{-8pt}

\subsection{Experimental Setup}
We evaluate AgentSoV on BFCL
\citep{liu2026agenthallu,patil2025berkeley}, Magentic-One
\citep{liu2026agenthallu,pmlr-v267-zhang25cq}, and AEGIS
\citep{kong2026aegis}. These datasets represent chain-structured tool use,
centralized orchestration, and DAG-structured information flow, respectively.
For failure attribution, we follow \citet{pmlr-v267-zhang25cq} and report step- and agent-level accuracy, together with Top-3 step accuracy and mean reciprocal rank (MRR).
The step- and agent-level accuracy are identical on BFCL because its
steps do not distinguish agent types. Because an AEGIS trajectory may contain multiple faulty agents,
we report exact-set accuracy and top-1 hit accuracy. 
Results are reported as mean $\pm$ standard deviation over five random seeds.
For each system-level analysis, AgentSoV is compared only with applicable baselines that provide sufficient implementation details and support the corresponding task.
We compare AgentSoV with six baselines spanning four categories: (1) Uniform Random, (2) three GPT-5.5
prompting strategies from \citet{pmlr-v267-zhang25cq}, (3) MASC
\citep{shen2026metacognitive}, and (4) OAT \citep{yeh2026tracing}. We evaluate two
variants of our method: AgentSoV-S uses a supervised residual-localization
head and step-level fault labels described in Appendix \ref{sec:macnet_agentsov}, whereas AgentSoV follows Section~\ref{sec:framework}.
Additional details are described in
Appendix~\ref{app_exp_set}.

\begin{table}[t]
\caption{Failure attribution performance on BFCL and Magentic-One.}
\label{tab:exp_attribution}
\centering
\resizebox{\linewidth}{!}{
\begin{tabular}{clcccc}
\toprule
\textbf{Dataset} & \textbf{Method} 
& \textbf{Step Accuracy}
& \textbf{Top-3 Step Accuracy}
& \textbf{MRR}
& \textbf{Agent Accuracy} \\
\midrule

\multirow{10}{*}{\footnotesize BFCL}
& Uniform Random & $11.7\%$ & $35.2\%$ & $31.7\%$ & $11.7\%$ \\
& GPT-5.5 (All-at-Once) & $16.5\%\pm1.2\%$ & $58.3\%\pm1.2\%$ & $39.4\%\pm1.4\%$ & $16.5\%\pm1.2\%$ \\
& GPT-5.5 (Step-by-Step) & $9.7\%\pm0.9\%$ & $49.8\%\pm2.8\%$ & $24.9\%\pm1.1\%$ & $9.7\%\pm0.9\%$ \\
& GPT-5.5 (Binary Search) & $2.9\%\pm1.1\%$ & $48.2\%\pm3.0\%$ & $23.7\%\pm1.4\%$  & $2.9\%\pm1.1\%$ \\
& MASC
    & $5.2\%\pm1.0\%$ & $33.8\%\pm1.4\%$& $29.5\%\pm0.4\%$& $5.2\%\pm1.0\%$ \\
& OAT
    & $8.4\%\pm0.5\%$ & $29.5\%\pm1.3\%$ & $27.9\%\pm0.3\%$  & $8.4\%\pm0.5\%$ \\
& AgentSoV-S ($F{=}0$)
    & $35.1\%\pm1.1\%$
    & $52.6\%\pm0.8\%$
    & $50.1\%\pm0.3\%$
    & $35.1\%\pm1.1\%$ \\
& AgentSoV-S
    & $36.9\%\pm0.6\%$
    & $\mathbf{64.3\%\pm2.2\%}$
    & $54.3\%\pm0.7\%$
    & $36.9\%\pm0.6\%$ \\
& AgentSoV ($F{=}0$)
    & $27.6\%\pm2.2\%$
    & $46.0\%\pm1.0\%$
    & $43.4\%\pm1.6\%$
    & $27.6\%\pm2.2\%$ \\
& AgentSoV
    & $\mathbf{37.5\%\pm2.0\%}$
    & $62.5\%\pm1.8\%$
    & $\mathbf{54.4\%\pm0.8\%}$
    & $\mathbf{37.5\%\pm2.0\%}$ \\
\midrule

\multirow{10}{*}{\footnotesize Magentic-One}
& Uniform Random & $13.6\%$ & $40.8\%$ & $34.4\%$ & $46.1\%$ \\
& GPT-5.5 (All-at-Once)
    & $31.5\%\pm1.9\%$  & $51.6\%\pm1.5\%$ & $42.8\%\pm0.8\%$ & $46.3\%\pm2.2\%$ \\
& GPT-5.5 (Step-by-Step)
    & $3.7\%\pm0.7\%$ & $46.9\%\pm1.7\%$ & $28.3\%\pm0.8\%$ & $51.9\%\pm3.1\%$ \\
& GPT-5.5 (Binary Search)
    & $5.6\%\pm1.4\%$ & $43.2\%\pm1.7\%$ & $20.8\%\pm0.5\%$ & $46.3\%\pm2.2\%$ \\
& MASC
    & $3.7\%\pm0.1\%$ & $28.1\%\pm2.2\%$& $26.0\%\pm2.0\%$ & $46.3\%\pm2.3\%$  \\
& OAT
    & $27.2\%\pm0.9\%$ & $54.1\%\pm0.7\%$ & $45.0\%\pm0.5\%$ & $53.1\%\pm1.7\%$ \\
& AgentSoV-S ($F{=}0$)
    & $25.2\%\pm2.8\%$ &$58.5\%\pm1.5\%$ & $46.4\%\pm1.0\%$ & $45.9\%\pm5.5\%$ \\
& AgentSoV-S
    & $36.3\%\pm3.6\%$ & $\mathbf{65.6\%\pm1.9\%}$ & $\mathbf{53.0\%\pm2.4\%}$ & $\mathbf{63.0\%\pm3.9\%}$ \\
& AgentSoV ($F{=}0$)
    & $35.9\%\pm2.2\%$
    & $52.6\%\pm0.9\%$
    & $45.6\%\pm1.12\%$
    & $61.9\%\pm1.5\%$ \\
& AgentSoV
    & $\mathbf{39.3\%\pm3.8\%}$
    & $54.1\%\pm1.4\%$
    & $47.7\%\pm2.1\%$
    & $61.9\%\pm4.2\%$ \\
\bottomrule
\end{tabular}
}
\end{table}

\subsection{Failure Attribution}
Tables~\ref{tab:exp_attribution} and~\ref{tab:aegis_injection_localization}
show that AgentSoV provides effective failure attribution across different MAS architectures. On BFCL, AgentSoV achieves the highest step accuracy ($37.5\%$). On Magentic-One, AgentSoV performs best for exact step localization ($39.3\%$), whereas AgentSoV-S achieves higher agent accuracy, top-3 accuracy, and MRR. This suggests that supervised residual-to-fault mappings are useful for identifying the responsible agent when multiple interactions from the same agent exhibit related failure patterns, while AgentSoV’s propagation model provides finer discrimination of the exact failure stage. On AEGIS, AgentSoV achieves $46.8\%$ exact-set and $89.7\%$ top-1 accuracy, outperforming AgentSoV-S, which indicates that explicit modeling of cross-agent propagation is beneficial when multiple faults interact through graph dependencies. Finally, the $\mF{=}0$ ablations generally reduce performance, with large drops for AgentSoV on BFCL and AgentSoV-S on Magentic-One, confirming that cross-stage propagation does provide crucial diagnostic information.

\begin{table}[htbp]
\centering
\caption{Failure attribution performance on AEGIS.}
\label{tab:aegis_injection_localization}
\resizebox{\linewidth}{!}{
\begin{tabular}{lcccc}
\toprule
\textbf{Method}
& \textbf{Exact-Set Accuracy}
& \textbf{Top-1 Accuracy}
& \textbf{Top-3 Accuracy}
& \textbf{MRR}
\\
\midrule
Uniform Random
& $9.0\%$ & $44.9\%$& $91.7\%$ &$67.6\%$ \\
GPT-5.5 (All-at-Once)
& $16.5\%\pm1.1\%$ & $84.2\%\pm0.8\%$ & $99.1\%\pm0.3\%$ & $91.5\%\pm0.3\%$  \\
GPT-5.5 (Step-by-Step)
& $8.3\%\pm2.2\%$ & $67.9\%\pm0.5\%$ 
& $99.8\%\pm0.3\%$ & $83.0\%\pm0.5\%$\\
GPT-5.5 (Binary Search)
& $12.2\%\pm2.6\%$ & $86.8\%\pm0.6\%$ & $99.6\%\pm0.3\%$ & $92.9\%\pm0.4\%$  \\
MASC
& $0.1\%\pm0.3\%$ & $76.7\%\pm2.6\%$ & $95.5\%\pm0.0\%$&$86.53\%\pm1.4\%$\\
OAT
& $4.9\%\pm1.1\%$ & $41.7\%\pm1.0\%$ &$87.5\%\pm1.2\%$ &$62.5\%\pm0.6\%$\\
AgentSoV-S ($F{=}0$)
& $33.1\%\pm0.9\%$ & $77.2\%\pm0.3\%$ &$98.1\%\pm0.0\%$ &$87.5\%\pm0.2\%$ \\
AgentSoV-S
& $33.6\%\pm2.1\%$ & $86.4\%\pm0.3\%$ &$99.4\%\pm0.0\%$ &$92.6\%\pm0.1\%$  \\
AgentSoV ($F{=}0$)
& $46.7\%\pm0.5\%$ & $88.6\%\pm0.3\%$ & $\mathbf{100.0\%\pm0.0\%}$ &$94.2\%\pm0.2\%$ \\
AgentSoV
& $\mathbf{46.8\%\pm0.41\%}$
& $\mathbf{89.7\%\pm0.0\%}$ & $\mathbf{100.0\%\pm0.0\%}$ &$\mathbf{94.9\%\pm0.0\%}$ \\
\bottomrule
\end{tabular}
}
\end{table}

We measure abnormal variation using the state-prediction error, $\|\vx_k-\hat{\vx}_k\|_2^2$, where $\hat{\vx}_k$ is defined in Eq.~\ref{eq:app_normal_prediction}. For each verified AEGIS repair, in which a non-faulty downstream receiver corrects a faulty parent's output, we compute the relative anomaly reduction from the parent to the receiver. A repair is detected when this reduction exceeds the specified threshold. Table~\ref{tab:macnet_repair} shows that AgentSoV achieves the highest recall at every threshold, detecting $74.7\%$ of repairs with any reduction. The results indicate that AgentSoV consistently captures substantial anomaly attenuation during downstream repair.

\begin{table}[htbp]
\centering
\caption{Recall of repaired AEGIS executions under different anomaly-reduction thresholds.}
\label{tab:macnet_repair}
\begin{tabular}{lccc}
\toprule
\textbf{Method}
& $>0\%$
& $\geq 10\%$
& $\geq 20\%$ \\
\midrule
Uniform Random
& $65.5\%$
& $45.1\%$
& $24.8\%$ \\
MASC
& $67.1\%$
& $67.1\%$
& $60.8\%$ \\
OAT
& $48.1\%$
& $21.5\%$
& $8.9\%$ \\
AgentSoV
& $\mathbf{74.7\%}$
& $\mathbf{68.4\%}$
& $\mathbf{63.3\%}$ \\
\bottomrule
\end{tabular}
\end{table}

Table~\ref{tab:bfcl_computation_cost} shows that AgentSoV has low one-time training costs and inference overhead. AgentSoV-S is the fastest method at inference ($0.0001$s), while AgentSoV requires $0.0242$s per trajectory, about $13.6\times$ faster than OAT and more than $110\times$ faster than the fastest GPT-5.5 strategy. MASC is faster than AgentSoV at inference but requires more training time and LLM encoding. AgentSoV avoids LLM inference and token consumption, yielding low costs without API latency and token usage. Additional results are in Appendix~\ref{app_add_res}.

\begin{table}[htbp]
\centering
\caption{Computational costs of BFCL.}
\label{tab:bfcl_computation_cost}
\begin{tabular}{@{}lccc@{}}
\toprule
\textbf{Method}
& \textbf{Training Time (s)}
& \textbf{Inference Time (s)}
& \textbf{LLM Tokens (In/Out)} \\
\midrule
GPT-5.5 (All-at-Once)   & 0     & 5.3427  & 618/440 \\
GPT-5.5 (Step-by-Step)  & 0     & 16.4219 & 2,586/1,026 \\
GPT-5.5 (Binary Search) & 0     & 2.6835  & 1,221/27 \\
MASC                    & 23.86 & 0.0054  & 5,102/0 \\
OAT                     & 7.82  & 0.3282  & 4,937/0 \\
AgentSoV-S              & 10.92 & 0.0001  & -- \\
AgentSoV                & 10.89 & 0.0242  & -- \\
\bottomrule
\end{tabular}%
\end{table}

\subsection{Stage-Input Influence}
\label{sec:exp_stage_input_sensitivity}
\begin{figure*}[htbp]
    \centering
    \begin{subfigure}[t]{0.225\textwidth}
        \centering
        \includegraphics[width=\linewidth]{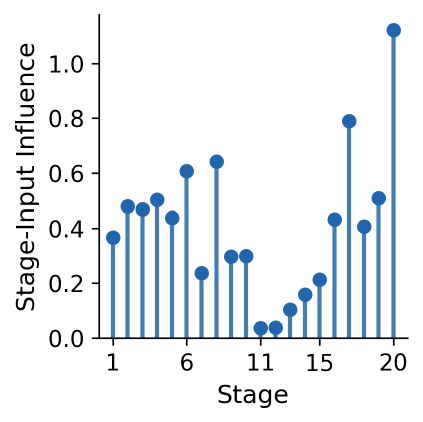}
        \caption{Input influence by stage.}
        \label{fig:magentic_sensitivity_stage}
    \end{subfigure}\hfill
    \begin{subfigure}[t]{0.225\textwidth}
        \centering
        \includegraphics[width=\linewidth]{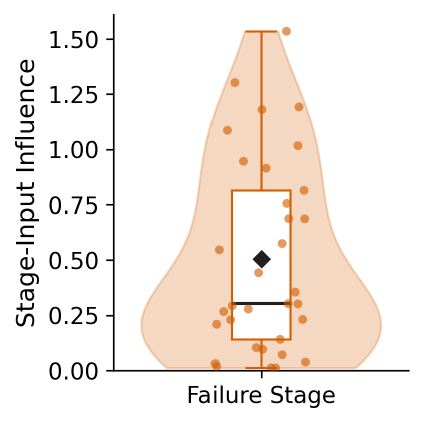}
        \caption{Failure-stage input influence.}
        \label{fig:magentic_sensitivity_failure}
    \end{subfigure}\hfill
    \begin{subfigure}[t]{0.225\textwidth}
        \centering
        \includegraphics[width=\linewidth]{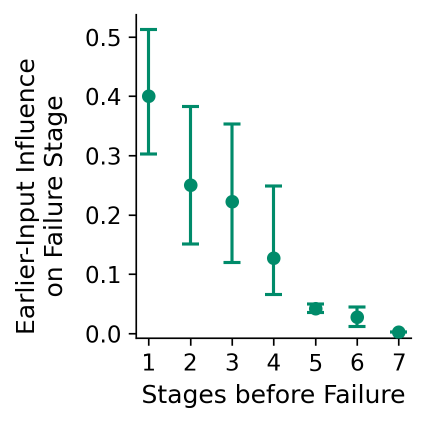}
        \caption{Influence of earlier inputs on failure stage.}
        \label{fig:magentic_sensitivity_predecessors}
    \end{subfigure}\hfill
    \begin{subfigure}[t]{0.28\textwidth}
        \centering
        \includegraphics[width=\linewidth]{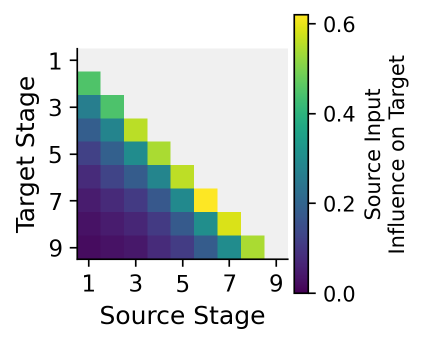}
        \caption{Source-to-target stage influence.}
\label{fig:magentic_sensitivity_matrix}
    \end{subfigure}
    \caption{Stage-input influence on Magentic-One.}
    \label{fig:magentic_sensitivity_four_panel}
\end{figure*}

Figure \ref{fig:magentic_sensitivity_four_panel} evaluates the stage-input influence as described in Eq.~\ref{eq:stage_to_stage_sensitivity}. Figure~\ref{fig:magentic_sensitivity_stage} shows substantial heterogeneity in how strongly individual stage inputs $\vu$ affect the final outcome $\vy_N$. Influence ranges from below \(0.1\) for several middle stages to around \(1.1\) at Stage 20, with other relatively influential stages such as Stages 6, 8, and 17 reaching roughly \(0.6\), \(0.65\), and \(0.8\), respectively. Thus, the learned dynamics do not assign comparable influence to all orchestrator instructions: some inputs produce little downstream change when removed, whereas others substantially alter the predicted final representation. Figure~\ref{fig:magentic_sensitivity_failure} shows a similarly heterogeneous distribution of failure-stage input influence on $\vy_N$. The mean influence is approximately \(0.5\), while many cases lie below \(0.3\) and a smaller number exceed \(1.0\), reaching about \(1.5\). This indicates that failure stages are not necessarily the stages whose inputs exert the strongest downstream influence.
Figure~\ref{fig:magentic_sensitivity_predecessors} further reveals a strong temporal decay in this influence. The input immediately preceding a failure stage has mean influence of approximately \(0.40\), which decreases to \(0.25\), \(0.22\), and \(0.13\) at two, three, and four stages earlier, respectively. In other words, the immediately preceding input is roughly three times as influential as an input four stages earlier. Figure~\ref{fig:magentic_sensitivity_matrix} shows the largest influence concentrate near the diagonal, where source and target stages are close, while entries become progressively weaker as their separation increases. 
Together, these results show that AgentSoV can identify non-uniform and predominantly local information propagation, revealing how stage inputs differ in downstream influence and how their effects attenuate across successive agent interactions.

\subsection{Virtual-Sensor Placement}
\begin{figure*}[htbp]
    \centering

    \begin{subfigure}[t]{0.245\textwidth}
        \centering
        \includegraphics[width=\linewidth]{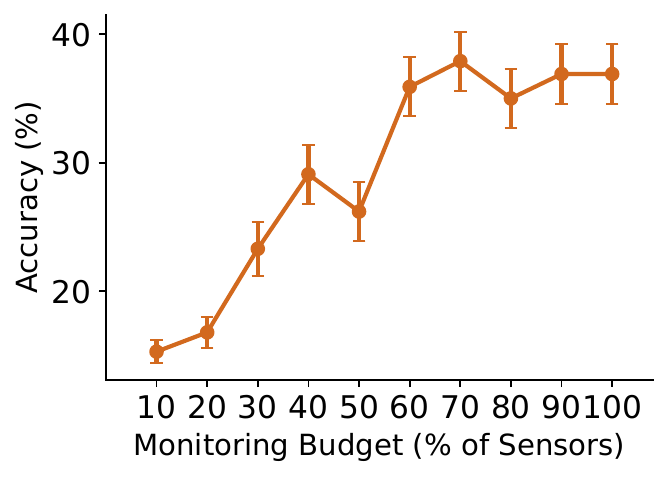}
        \caption{Sensor: BFCL.}
        \label{fig:sensor_bfcl_stage}
    \end{subfigure}
    \hfill
        \begin{subfigure}[t]{0.245\textwidth}
        \centering
        \includegraphics[width=\linewidth]{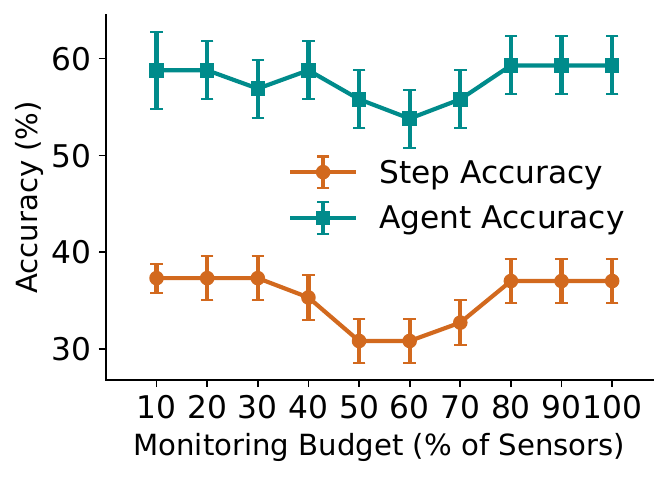}
        \caption{Sensor: MO.}
        \label{fig:sensor_magentic_stage}
    \end{subfigure}
    \hfill
    \begin{subfigure}[t]{0.245\textwidth}
        \centering
        \includegraphics[width=\linewidth]{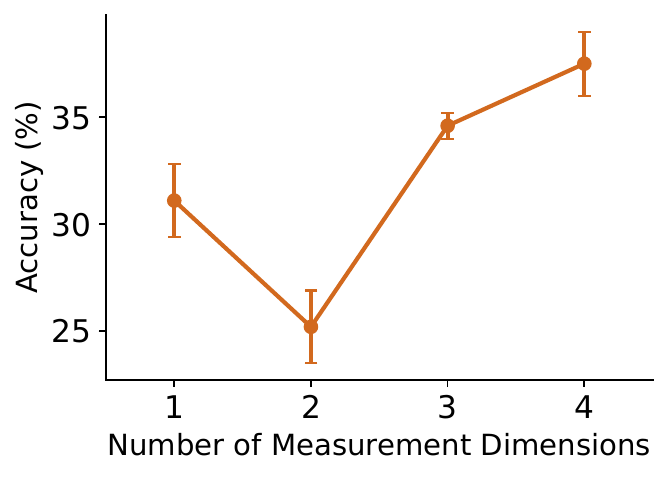}
        \caption{$\vy$ dimensions: BFCL.}
        \label{fig:sensor_bfcl_coord}
    \end{subfigure}
    \hfill
    \begin{subfigure}[t]{0.245\textwidth}
        \centering
        \includegraphics[width=\linewidth]{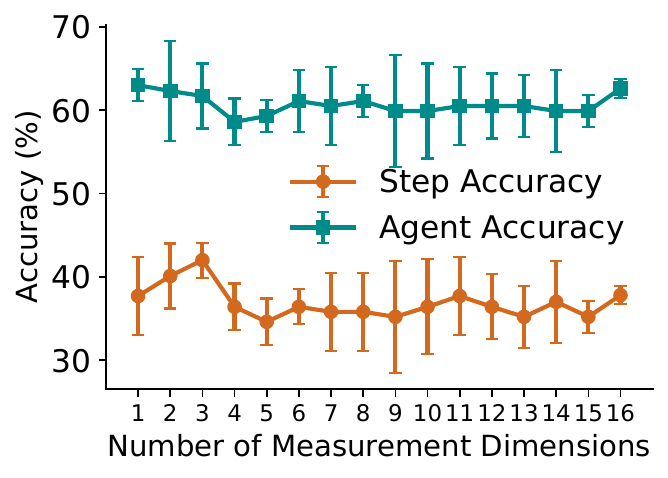}
        \caption{$\vy$ dimensions: MO.}
        \label{fig:sensor_magentic_coord}
    \end{subfigure}
    \caption{Virtual-sensor placement and $\vy$-dimension selection under different observation budgets: (a \& c) BFCL and (b \& d) Magentic-One (MO).
    }
    \label{fig:sensor_selection}
\end{figure*}

Figure~\ref{fig:sensor_selection} evaluates virtual-sensor placement and $\vy$-dimension selection on BFCL and Magentic-One. Figures~\ref{fig:sensor_bfcl_stage} and~\ref{fig:sensor_magentic_stage}  vary the percentage of monitored stages following Section~\ref{sec:sensor_placement}, while Figures~\ref{fig:sensor_bfcl_coord} and~\ref{fig:sensor_magentic_coord}  similarly vary the monitored dimensions in each $\vy_k$. BFCL benefits from broader stage coverage and all four dimensions, whereas Magentic-One matches full-observation performance with only $10\%$ of stages and achieves its highest step accuracy with three dimensions, indicating substantial measurement redundancy. These results show that AgentSoV can identify compact monitoring configurations and reveal when additional measurements are necessary.


\begin{figure}[htbp]
\centering

\begin{subfigure}[t]{0.3\columnwidth}
    \centering
    \includegraphics[width=\linewidth]{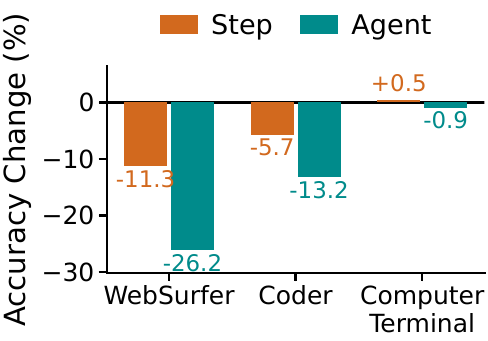}
    \caption{Magentic-One.}
    \label{fig:hidden_agent_magentic}
\end{subfigure}
\hspace{0.05\textwidth}
\begin{subfigure}[t]{0.3\columnwidth}
    \centering
    \includegraphics[width=\linewidth]{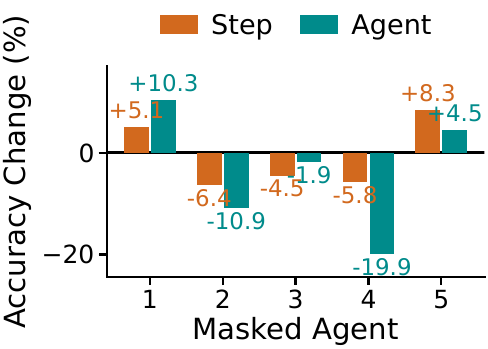}
    \caption{AEGIS.}
    \label{fig:hidden_agent_aegis}
\end{subfigure}

\caption{Agent-masking effects on failure-attribution accuracy relative to full observation.}
\label{fig:hidden_agent_ablation}
\end{figure}

Figure~\ref{fig:hidden_agent_ablation} evaluates agent-specific effects by masking all call-node $\vy$ from one agent type. Masking WebSurfer on Magentic-One and Agent~4 on AEGIS causes the largest accuracy losses, showing that their outputs provide important diagnostic evidence. In contrast, masking AEGIS Agents~1 and~5 improves attribution, showing that their $\vy$ obscure fault signatures. Diagnostic value depends on an agent's role and position, and additional monitoring does not always improve attribution.

\section{Conclusion}
We introduced AgentSoV, a virtual-sensing framework that represents heterogeneous LLM MAS trajectories as diagnostic stages in a stochastic dynamic system. Its architecture-aware yet unified model captures cross-stage variation in chain-structured, orchestration-based, and DAG-structured systems, supporting failure attribution, stage-input influence, and virtual-sensor placement without MAS replay. Experiments on BFCL, Magentic-One, and AEGIS show improved failure localization, differentiated downstream influence across stage inputs, and preserved diagnostic performance with fewer measurements. Future work can extend AgentSoV beyond linear state-space dynamics and automate stage and measurement construction.

\bibliography{iclr2027_conference}
\bibliographystyle{iclr2027_conference}

\newpage

\appendix

\section{Related Work}\label{app_rel_work}

\subsection{SoV}
SoV was a branch of state-space models developed for multistage manufacturing systems to model how variation introduced at one process stage propagates to and is
observed at downstream stages \citep{shi2006stream, ceglarek2004time}. Early work formulated this propagation using knowledge-based state-space models, enabling process-level diagnosis of variation sources across multiple stations \citep{huang2000stream}. Subsequent studies connected the propagation model to system diagnosability, analogous to observability in control systems, and examined whether different variation sources can be distinguished from available measurements \citep{ding2002diagnosability,zhou2003diagnosability}. The same structure also provides a principled basis for sensor placement: measurements can be allocated across stages according to how variation is
transmitted and detected, rather than treating each station independently \citep{liu2005optimal, bastani2016compressive}. AgentSoV extends this SoV perspective to address the challenges in LLM MASs. LLM MASs often fail because local errors (such as faulty specifications, misaligned messages, or invalid intermediate results) propagate downstream, causing symptoms far from their source and evading sparse or output-only checks. AgentSoV explicitly models this propagation by constructing virtual measurements from observable execution signals and modeling how task-relevant variation propagates across diagnostic stages, enabling fault attribution, identification of redundant stage contributions, and cost-aware placement of monitors where they provide the greatest diagnostic value.

\subsection{Failure Attribution in LLM MASs} 

Failure attribution for MAS has been approached through four main methodologies, including LLM-based trajectory judging and decomposition, supervised attribution models, causal or counterfactual intervention, and structure-based methods such as spectrum analysis and anomaly detection. 

The widely adopted Who\&When benchmark \citep{pmlr-v267-zhang25cq} formalizes agent and step level attribution for single agent and decisive step. AEGIS scales supervision through synthetically injected and labeled errors \citep{kong2026aegis}. These formulations may not capture naturally occurring failures with multiple agents and multiple interacting causes, and their dependence on failure annotations limits transfer to new systems. 
AgentSoV complements these benchmarks with an explicit model of cross-stage variation propagation, using residual evidence to localize abnormal sources.

On the representative methods, SDBL reduces the reasoning burden by separating failure-scope detection from step localization \citep{sun2026scope}, and AgenTracer trains a specialized model using counterfactual-replay supervision \citep{zhang2026agentracer}. However, these methods remain tied to trajectory representations and training distributions whose semantics change across sequential, orchestrated, and DAG-based architectures. AgentSoV relocates this architecture-specific adaptation into explicit definitions of diagnostic stages, inputs, states, and measurements, after which attribution is performed through a common state-space formulation.

The intervention-based methods, on the other hand, provide stronger causal evidence. DoVer tests candidate explanations by modifying executions \citep{ma2026dover}, while FAMAS estimates suspiciousness from repeated trajectory replay \citep{ge2026spectrum}. Their principal limitation is operational cost and the need for a re-executable system and environment. AgentSoV compares an observed trajectory with its normal model prediction and attribute the resulting residual without replay or per-trace LLM inferences in MAS.

Finally, OAT learns the dynamics of successful trajectories and detects deviations without failure labels, but primarily returns anomaly scores rather than an explicit account of where and how abnormal variation entered \citep{yeh2026tracing}. AgentSoV extends this dynamical perspective by decomposing residuals into stage-specific disturbance directions and propagating their signatures across downstream observations, producing more interpretable attribution while exposing when candidate failure sources are observationally difficult to distinguish.

\subsection {Information Analysis in LLM MASs} 

The information analysis is primarily studied through communication pruning, topology optimization, heuristic influence scoring, and cooperative credit assignment. AgentPrune \citep{zhang2024cut} and AgentDropout \citep{wang2025agentdropout} learn masks that remove redundant communication edges or agents while preserving task performance. Sparse-debate and topology-generation methods optimize communication structures using task-level reward \citep{li2024multi,zhang2026radar,li2026assemble}. These methods demonstrate that substantial communication can be omitted, but their learned structures provide limited insight into why a particular message was redundant or not. Also, reported gains can depend strongly on the initial topology, prompts, and token-cost accounting \citep{li2026discovering,zhang2026evaluation}. AgentSoV addresses this interpretability gap by representing each message or stage input, and propagating the resulting counterfactual change through an identified dynamical model to measure its effect on final output.

Moreover, DyLAN directly estimates individual-agent importance from downstream effects \citep{liu2024a}, while causal analyses examine how correct and erroneous information spreads through alternative communication topologies \citep{shen2025information}. However, DyLAN relies on an unsupervised propagation heuristic, and topology-level analyses do not isolate the contribution of a specific stage input. AgentSoV instead produces a stage-level influence score, that distinguishes information while accounting for its modeled downstream propagation.

As a principled alternative, Shapley-based methods distribute credit across agents or messages, but evaluate many possible coalitions grows exponentially with system size \citep{hua2025shapley}. Under its linear approximation, AgentSoV computes each counterfactual contribution in closed form from the fitted model, avoiding coalition enumeration and repeated MAS execution.

Evaluation remains an open challenge: existing studies typically combine general capability benchmarks such as MMLU, GSM8K, MATH, and HumanEval with token or API-cost measurements, rather than using a dedicated benchmark for information contribution \citep{yang2026compared,chang2025jopa}. AgentSoV offers a common model-based quantity that could support such evaluation across architectures, although its fault-free reference inputs and one-at-a-time perturbations must be carefully calibrated because message content and stage inputs may interact.

\subsection{Monitoring for MASs} 

The runtime monitoring is currently addressed through observability frameworks, proactive guardrails, selective verification, and trace-sampling methods. AgentOps catalogs the artifacts that agents should emit for debugging \citep{dong2024agentops}, but treats monitoring locations as predetermined and does not determine which observations are necessary for fault attribution. Runtime guardrails such as ProbGuard and AGrail predict unsafe states or adapt safety checks over time \citep{wang2025probguard,luo2025agrail}, while ToolSafe uses TS-Bench to evaluate step-level tool-invocation monitoring \citep{mou2026toolsafe}. Although these approaches improve detection, their checks are generally attached to fixed locations or applied at every relevant step, which can be expensive when intermediate evaluations require LLM judges or process supervision. AgentSoV reduces this cost by selecting a budgeted subset of stages whose measurements distinguish the largest range of possible disturbance sources.

Weak-to-strong monitoring studies on SHADE-Arena and CUA-SHADE-Arena further show that monitor scaffolding, such as where and how observations are collected, can matter more than monitor-model strength \citep{kale2026reliable}. However, these benchmarks evaluate robustness of a given monitoring configuration rather than optimizing the configuration itself. AgentSoV formalizes this missing selection problem using stage-specific disturbance signatures and a log-determinant objective that favors measurements that are both informative and non-redundant. 

SIEVE similarly avoids exhaustive verification by escalating only when a heuristic suspicion signal is triggered \citep{liang2025sieve}, but such triggers may overlook failures that remain locally inconspicuous while propagating downstream. Drawing on established optimal sensor-placement and Bayesian D-optimal design methods \citep{guestrin2005near,huan2024optimal}, AgentSoV adapts this principle to stage-level MAS monitoring. Because no dedicated benchmark currently evaluates budgeted monitor placement in MAS, AgentSoV’s partial-observation experiments also provide a foundation for measuring how much attribution capability is retained as monitoring cost decreases.

\section{Instantiating AgentSoV Across MAS Architectures}
\label{sec:adaptation}

The preceding AgentSoV formulation provides a common interface, i.e., diagnostic stages
$s_k$, inputs $\vu_k$, states $\vx_k$, virtual measurements $\vy_k$, disturbances
$\vz_k$, and cross-stage propagation, without prescribing a single
trajectory decomposition for every MAS. Different execution architectures
instantiate these quantities differently. We illustrate this with three
representative settings: chain-structured tool use, centralized orchestration,
and DAG-structured information flow.

\subsection{Chain-Structured MAS}
\label{sec:tool_calling}

In a chain-structured system, the basic dependency in
Eq.~\ref{eq:state_equation_s3} applies directly: each stage primarily inherits
variation from the immediately preceding stage. We instantiate this setting
using BFCL data \citep{patil2025berkeley,liu2026agenthallu}, where an
assistant repeatedly issues tool calls and incorporates the resulting tool
responses over multiple turns.

Each diagnostic stage $s_k$ corresponds to one assistant turn together with
its tool calls and returned results. The process input summarizes the action
issued before observing those results:
\begin{equation*}
\vu_k =
\begin{bmatrix}
n_k^{\mathrm{call}} &
n_k^{\mathrm{arg}} &
n_k^{\mathrm{write}} &
\rho_k
\end{bmatrix}^{\!\top}
\in\R^{4},
\end{equation*}
where $n_k^{\mathrm{call}}$, $n_k^{\mathrm{arg}}$, and
$n_k^{\mathrm{write}}$ denote the numbers of tool calls, supplied arguments,
and environment-modifying calls, respectively, and $\rho_k$ is the fraction
of calls repeated from earlier stages. All four quantities are calculated
directly from the execution trace.

BFCL contains four corresponding failure types: incorrect tool arguments,
missing required calls, unnecessary tool calls, and parallel-call conflicts.
Because the measurement coordinates and failure mechanisms have explicit
semantics, we choose to configure
\begin{equation*}
\begin{aligned}
\vy_k =
\begin{bmatrix}
y_k^{\mathrm{schema}} &
y_k^{\mathrm{inaction}} &
y_k^{\mathrm{exec}} &
y_k^{\mathrm{accum}}
\end{bmatrix}^{\!\top},\\
\mC =
\begin{bmatrix}
1 & 0 & 0 & 0 \\
0 & 1 & 0 & 0 \\
0 & 0 & 1 & 0 \\
0 & 0 & 0.5 & 1
\end{bmatrix},
\quad
\mH =
\begin{bmatrix}
1 & 0 & 0 & 0 \\
0 & 1 & -1 & 0 \\
0 & 0 & 0.5 & 1 \\
0 & 0 & 0 & 1
\end{bmatrix}.
\label{eq:bfcl_CH}
\end{aligned}
\end{equation*}
The first three rows of $\mC$ directly observe their corresponding state
coordinates, while its final row allows the accumulated-error measurement to
reflect both current execution error and previously accumulated variation.
The four columns of $\mH$ correspond, respectively, to the four failure types.
For example, an unnecessary call decreases the inaction component while
increasing execution-error variation, whereas a parallel-call conflict affects
both current and accumulated execution errors.

Tool-calling trajectories provide structured execution signals, so following the definition of the $\mC$ and $\vy$, the four interpretable diagnostic coordinates can be written as:
\begin{equation*}
\vx_k =
\begin{bmatrix}
x_k^{\mathrm{schema}} &
x_k^{\mathrm{inaction}} &
x_k^{\mathrm{exec}} &
x_k^{\mathrm{accum}}
\end{bmatrix}^{\!\top}.
\label{eq:bfcl_state_measurement2}
\end{equation*}
These coordinates represent schema defects, inaction, current execution
errors, and accumulated unresolved errors. Specifically,
$y_k^{\mathrm{schema}}$ is the fraction of argument keys at stage $k$ that
occur only once in the trajectory, $y_k^{\mathrm{inaction}}$ is one if no
tool call is made and zero otherwise, $y_k^{\mathrm{exec}}$ counts failed
tool calls, and $y_k^{\mathrm{accum}}$ counts unresolved execution errors
accumulated through stage $k$.

No change to the general state-space logic is required. The learned $\mF$
characterizes how diagnostic variation persists across tool-use turns, while
$\mB$ characterizes how the current tool-use pattern changes the diagnostic
state. Thus, the BFCL instantiation differs from the running textual example
primarily in how $\vu_k$, $\vx_k$, $\vy_k$, $\mC$, and $\mH$ are materialized from a
structured execution log.

\subsection{Orchestration-Based MAS}
\label{sec:magentic_agentsov}

An orchestration-based MAS introduces a dependency that is absent from a
simple chain: persistent task-level information maintained by a central
controller may influence many downstream stages. Magentic-One
\citep{fourney2024magentic,liu2026agenthallu} provides a representative
example and closely resembles the Orchestrator-WebSurfer-Coder system used
throughout our illustration.

A Magentic-One execution contains an initial orchestrator fact sheet, repeated
orchestrator-agent interactions, and a final response. We therefore map the
trajectory as
\begin{equation*}
\underbrace{\tau_1}_{s_0}
\rightarrow
\underbrace{(\tau_2,\tau_3)}_{s_1}
\rightarrow \cdots \rightarrow
\underbrace{(\tau_{2K},\tau_{2K+1})}_{s_K}
\rightarrow
\underbrace{\tau_{2K+2}}_{s_{K+1}},
\label{eq:magentic_parse}
\end{equation*}
where $s_0$ represents the persistent orchestration information,
$s_1,\ldots,s_K$ are orchestrator-worker stages, and $s_{K+1}$ represents
the delivered response. Relative to
Eq.~\ref{eq:example_stages}, this simply retains the individual
orchestrator-worker interactions rather than aggregating them into a smaller
number of functional stages.

For an execution stage $s_k (k\geq 1)$, the information dispatched by the orchestrator
forms the process input, while the corresponding worker response provides the
virtual measurement. In our Magentic-One instantiation,
\begin{equation*}
\vu_k = \mP\,\mathcal{E}(\tau_{2k}),
\qquad
\vy_k = \mP\,\mathcal{E}(\tau_{2k+1}),
\label{eq:magentic_vars}
\end{equation*}
where $\mathcal{E}(\cdot)$ is a frozen sentence encoder and $\mP$ is a PCA
projection fitted using the training split. Thus, the numerical
representations differ from those used in BFCL, but $\vu_k$ and $\vy_k$ retain
exactly the roles defined by
Eqs.~\ref{eq:state_equation_s3}-\ref{eq:observation_equation_s3}.

The initial orchestration stage requires separate treatment because its
information is available to many later stages rather than only to $s_1$.
Magentic-One's fact sheet explicitly separates verified facts, information to
derive or guess, and planned actions. We therefore construct
\begin{equation*}
\vy_0 =
\begin{bmatrix}
n_{\mathrm{verified}} &
n_{\mathrm{derive/guess}} &
n_{\mathrm{action}}
\end{bmatrix}^{\!\top},
\label{eq:plan_measurement}
\end{equation*}
where the three terms summarize the plan's grounding, unresolved information,
and scope. The corresponding orchestration state $\vx_0$ is retained as a
shared upstream state for the downstream execution.

The architectural change can then be represented by adding one term to the
basic state equation:
\begin{equation*}
\vx_k
=
\mF \vx_{k-1}
+
\mB \vu_k
+
\mG\vx_0
+
\mH\vz_k
+
\vw_k,
\qquad
\vy_k=\mC\vx_k+\vv_k,
\quad k=1,\ldots,K,
\label{eq:magentic_sov}
\end{equation*}
where $\mG$ captures the direct influence of the persistent orchestration state
on each downstream stage. The term $\mG_k \vx_0$ admits two equivalent reformulations to the original formula: augmenting the state as $\tilde{\vx}_k = [\vx_k^\top, \vx_0^\top]^\top$ with transition
$\begin{bmatrix} 
\mF_k & \mG_k \\ 
0 & \mI 
\end{bmatrix}$,
or appending $\vx_0$ to each $\vu_k$. We maintain this term explicitly for two reasons. First, in orchestrator-worker designs, the task specification and global plan are re-supplied to every worker rather than relayed sequentially by preceding stages; thus, $\vx_0$ acts as an exogenous input across all stages. Second, separating these channels preserves the interpretation of $\mB_k \vu_k$ as stage-specific control information, which is necessary for the intervention analysis in Section~\ref{sec:information_requirement}. Folding $\vx_0$ directly into $\vu_k$ would mean that ``removing a stage input'' inadvertently wipes out the global task definition.

This decomposition separates two qualitatively different propagation
mechanisms. Variation in $\vx_0$, such as an incorrect global plan or poorly
grounded fact sheet, acts as a common upstream source and can influence many
later stages through $\mG$. By contrast, abnormal variation introduced by one
worker enters through $\mH\vz_k$ and subsequently propagates to later stages
through $\mF$. Consequently, an incorrect plan and an incorrect intermediate
agent response can produce different cross-stage signatures even when they
eventually yield the same wrong final answer. If an orchestration-based MAS
has no explicit persistent planning state, the $\mG\vx_0$ term can simply be
omitted, recovering the basic formulation.

\subsection{DAG-Structured MAS}
\label{sec:macnet_agentsov}
A DAG-structured MAS differs from the previous two cases because a stage may
depend on several upstream stages rather than on a unique predecessor. We
instantiate this setting on AEGIS \citep{kong2026aegis}, where each trajectory
contains ten LLM calls connected by a trajectory-specific directed acyclic
graph. Each call forms one diagnostic stage, and an edge from stage $j$ to
stage $k$ exists when the response produced at $j$ is included in the input
to $k$.

For each call, the process input $\vu_k\in\R^{16}$ is obtained by
encoding the complete input-message list with a frozen encoder and projecting
it to 16 dimensions using PCA. The response is represented analogously to
construct $\vx_k\in\R^{16}$. PCA is fitted using non-injected training
calls. Because the response representation is itself directly observed, this
instantiation uses $\vy_k=\vx_k, \mC_k=\mI.$

The only substantive change to Eq.~\ref{eq:state_equation_s3} is the
cross-stage propagation term. Let $\mathrm{pa}(k)$ denote the set of
diagnostic stages whose outputs directly enter stage $k$. Instead of receiving
propagated variation from one preceding stage, stage $k$ aggregates it across
its parents:
\begin{equation}
\vx_k
=
\mB \vu_k
+
\frac{1}{|\mathrm{pa}(k)|}
\sum_{j\in\mathrm{pa}(k)}
\mF \vx_j
+
\vw_k,
\label{eq:macnet_sov}
\end{equation}
with an intercept included in the implementation through the usual augmented
input representation. The term $\mB\vu_k$ captures the call-local relationship
between the complete information supplied to the current call and its
response, while $\mF\vx_j$ captures variation inherited from an upstream call.
The parameters $\mF$ and $\mB$ are shared across call positions and trajectories
and are fitted using non-injected training calls.

The formulation above defines the normal DAG dynamics used by AgentSoV.
Because AEGIS also provides injected-fault labels, we use the same dynamics to
construct AgentSoV-S and examine whether a supervised residual-to-fault mapping
improves localization. AgentSoV-S fits a local predictor using only
$\mB\vu_k$ and a graph-aware predictor using the full dynamics in
Eq.~\ref{eq:macnet_sov}, both on non-injected training calls. For each call, it
combines the standardized local residual
$\vx_k-\hat{\vx}_k^{\mathrm{local}}$ with the graph-explained residual difference
$\hat{\vx}_k^{\mathrm{full}}-\hat{\vx}_k^{\mathrm{local}}$, and a linear classifier
shared across call positions produces the call-level fault probability. Agent
scores are obtained by taking the maximum probability over their calls, after
which a validation-selected decoding rule predicts the faulty-agent set. Thus,
AgentSoV-S retains the graph-aware representation of AgentSoV but learns how to
map its residual evidence to fault predictions.

For both variants, Eq.~\ref{eq:macnet_sov} is the DAG analogue of the
$\mF_k\vx_{k-1}$ term in Eq.~\ref{eq:state_equation_s3}, rather than a separate
diagnostic model. For a chain there is one propagation path from an upstream
stage to a downstream stage; for a DAG, variation may reach the same
downstream stage through several directed paths. Accordingly, the
multi-stage propagation quantities used in Section~\ref{sec:diagnosis} are
obtained by aggregating the effects of all directed paths connecting the
source and observation stages. The meanings of input, state, measurement,
and disturbance remain unchanged.

Taken together, these three instantiations illustrate the distinction between
the architecture-aware construction and the
architecture-independent analytical language of AgentSoV. A
tool-calling chain, an orchestrated system, and a DAG expose different
execution structures and therefore require different stage mappings and
dependency patterns. Once represented, however, the same quantities retain
the same interpretation: $\vu_k$ captures information entering a stage,
$\vx_k$ captures task-relevant variation carried through the execution,
$\vy_k$ captures what is virtually observed, $\mH_k\vz_k$ represents abnormal
variation when modeled explicitly, and the transition terms describe how
upstream variation reaches downstream stages. The analyses in
Section~\ref{sec:diagnosis} build directly on this common representation.

\section{Computational Details for Residual-Based Failure Attribution}
\label{app:residual_diagnosis}

Section~\ref{sec:fault_attribution} describes failure attribution in terms of
the propagated disturbance signature 
$\mG_{\ell,k}$,
which characterizes how abnormal variation introduced at stage $k$ becomes
observable at a downstream stage $\ell$. This appendix gives the corresponding
residual-based computational formulation used to identify the source of
abnormal variation from an observed execution.

\subsection{Normal Prediction and Diagnostic Residuals}
\label{app:diagnostic_residuals}

We first define the execution expected under the fitted normal dynamics.
Setting $\vz_k=\mathbf{0}$ in Eq.~\ref{eq:state_equation_s3}, the one-step state and
measurement predictions are
\begin{align}
    \hat{\vx}_k &= \mF_k\hat{\vx}_{k-1}+\mB_k\vu_k, 
    \label{eq:app_normal_prediction}\\
    \hat{\vy}_k &= \mC_k\hat{\vx}_k,
\end{align}
where $\hat{\vx}_k$ and $\hat{\vy}_k$ denote the state and virtual measurement
predicted in the absence of an explicit abnormal disturbance. For an observed
execution, we define the stage-wise diagnostic residual
\begin{equation}
    \ve_k
    =
    \vy_k-\hat{\vy}_k.
    \label{eq:app_residual}
\end{equation}
Thus, $\ve_k$ measures the part of the observed virtual measurement that is not
explained by the fitted normal AgentSoV dynamics.

Importantly, a nonzero residual by itself does not identify where abnormal
variation originated. The residual at stage $k$ may reflect a disturbance
introduced locally, variation propagated from an earlier stage, ordinary
process and measurement noise, or model mismatch. The state-space structure
allows these possibilities to be separated through their different
cross-stage patterns.

Suppose an abnormal disturbance $\vz_k$ is introduced at stage $k$, with no
additional abnormal disturbance introduced elsewhere. As shown in
Section~\ref{sec:fault_attribution}, its contribution to the virtual
measurement at a downstream stage $\ell\geq k$ is
$\mG_{\ell,k}\vz_k$. By causality, the same disturbance cannot affect measurements
at stages preceding $k$. The diagnostic residual therefore decomposes as
\begin{equation}
\ve_\ell
=
\begin{cases}
    \boldsymbol{\epsilon}_\ell,
        & \ell<k,\\[2mm]
    \mG_{\ell,k}\vz_k+\boldsymbol{\epsilon}_\ell,
        & \ell\geq k,
\end{cases}
\label{eq:app_residual_decomposition}
\end{equation}
where $\boldsymbol{\epsilon}_\ell$ collects the residual variation expected under normal
execution, including propagated process noise, measurement noise, and model
mismatch.

Equation~\ref{eq:app_residual_decomposition} provides the key diagnostic
principle. A disturbance is not identified merely because stage $k$ has a
large residual. Rather, a candidate source at stage $k$ is plausible when the
residuals observed from $k$ onward are consistent with the downstream pattern
predicted by $\mG_{\ell,k}\vz_k$.

For example, in the Orchestrator-WebSurfer-Coder system, an abnormal
retrieval disturbance at $s_1$ can generate residuals at the retrieval,
reasoning, and final-response stages. In contrast, a disturbance introduced by
the Coder at $s_2$ cannot explain an abnormal residual already present at
$s_1$. This temporal and structural constraint is what distinguishes
AgentSoV attribution from independently scoring each stage for anomaly.

\subsection{Trajectory-Level Residual Representation}
\label{app:trajectory_residual}

To use the residual pattern jointly across the execution, we stack the
stage-wise residuals into
\begin{equation*}
    \ve
    =
    \begin{bmatrix}
        \ve_1^\top &
        \ve_2^\top &
        \cdots &
        \ve_N^\top
    \end{bmatrix}^{\!\top}.
    \label{eq:app_stacked_residual}
\end{equation*}
For a candidate disturbance source at stage $k$, we similarly collect its
predicted effects over the complete trajectory into the trajectory-level disturbance signature
\begin{equation*}
    \mG_k
    =
    \begin{bmatrix}
        \mathbf{0} &
        \cdots &
        \mathbf{0} &
        \mG_{k,k}^{\top} &
        \mG_{k+1,k}^{\top} &
        \cdots &
        \mG_{N,k}^{\top}
    \end{bmatrix}^{\!\top}.
    \label{eq:app_trajectory_signature}
\end{equation*}
The leading zero blocks encode causality: abnormal variation introduced at
stage $k$ cannot influence observations made before that stage. The remaining
blocks describe how that same disturbance should appear throughout its
downstream execution. Let $\boldsymbol{\epsilon}
=
\begin{bmatrix}
    \boldsymbol{\epsilon}_1^\top &
    \cdots &
    \boldsymbol{\epsilon}_N^\top
\end{bmatrix}^{\!\top}$.
Then Eq.~\ref{eq:app_residual_decomposition} becomes
\begin{equation}
    \ve
    =
    \mG_k \vz_k
    +
    \boldsymbol{\epsilon}
    \label{eq:app_single_source}
\end{equation}
when abnormal variation originates from a single stage.

This representation makes explicit why the downstream structure is useful for
diagnosis. Two candidate stages may both exhibit large local residuals, but
their trajectory-level signatures $\mG_k$ generally differ because
they enter the system at different locations and propagate through different
sequences of $\mF_\ell$ matrices before becoming visible through the
corresponding $\mC_\ell$ matrices.

To account for the different scales and correlations of residual components,
we estimate the covariance of normal residuals from fault-free training
trajectories. If the stage-wise residual covariance is denoted by $\mSigma$,
a simple implementation assumes a common covariance across stages and uses $\mSigma_{\ve}=\mI_N\otimes\mSigma.$
More general empirical covariance models can be used when sufficient
fault-free trajectories are available. In either case, the covariance serves
to discount residual directions that commonly fluctuate during normal
execution and to give greater weight to deviations that are unusual relative
to the fitted normal dynamics.

\subsection{Joint Attribution with Sparse Disturbances}
\label{app:sparse_attribution}

Although some benchmarks assume a single responsible stage, an MAS execution
may contain abnormal variation introduced at multiple stages. The residual
model then generalizes naturally to $\ve
    =
    \sum_{k=1}^{N}
    \mG_k \vz_k
    +
    \boldsymbol{\epsilon}.$
Each $\vz_k$ is a vector associated with one candidate source stage, while
$\mG_k\vz_k$ is the complete downstream residual pattern predicted from
that source.

Because failures are typically expected to originate from only a small subset
of the diagnostic stages, we estimate the candidate disturbances using a
group-sparse objective:
\begin{equation}
\{\hat{\vz}_k\}_{k=1}^{N}
=
\arg\min_{\{\vz_k\}}
\left\|
    \ve
    -
    \sum_{k=1}^{N}\mG_k\vz_k
\right\|_{\mSigma_{\ve}^{-1}}^{2}
+
\lambda
\sum_{k=1}^{N}
\|\vz_k\|_2 ,
\label{eq:app_sparse_attribution}
\end{equation}
where $\|\va\|_{\mSigma_{\ve}^{-1}}^2
=
\va^\top\mSigma_{\ve}^{-1}\va$ and $\lambda>0$ controls the sparsity of the attribution.

The first term in Eq.~\ref{eq:app_sparse_attribution} measures how well the
candidate disturbances jointly explain the observed trajectory-wide residual,
after accounting for residual variation expected during normal execution.
The second term is an $\ell_2/\ell_1$ group penalty: the $\ell_2$ norm treats
all components of the disturbance at one stage as a group, while the outer
sum encourages entire stage-level disturbance vectors to be zero. Hence,
stages with $\hat{\vz}_k\neq \mathbf{0}$ are selected as candidate sources of abnormal
variation.

This formulation is useful because attribution is performed through
explanation of the complete residual trajectory. A stage is selected only when the disturbance entering through its $\mH_k$ direction and propagating through the learned $\mF$ dynamics explains the observed downstream deviations.
A stage with a large local residual but an incompatible downstream signature
can therefore receive less attribution than an earlier stage whose propagated
effect explains several later residuals simultaneously. When a benchmark requires a single responsible stage, the group-sparse
solution can be reduced to a stage-level ranking using $\hat{k}=\arg\max_{k}\|\hat{\vz}_k\|_2.$ When multiple causes are allowed, the nonzero groups themselves provide the multi-stage attribution.

\subsection{Relationship to the Analyses in Section~\ref{sec:diagnosis}}
\label{app:residual_relationship}

The residual formulation above is one computational realization of the
system-level ideas in Section~\ref{sec:diagnosis}. Failure attribution uses
the disturbance channel $\mH_k\vz_k$ and asks which propagated disturbance
signatures best explain the observed residuals. Stage-input influence instead perturbs the process-input channel $\mB_k\vu_k$ and propagates
that change through the same learned transition matrices. Virtual-sensor
placement asks which observation matrices $\mC_\ell$, and therefore which
virtual measurements $\vy_\ell$, make the candidate propagated disturbance
signatures most distinguishable relative to normal residual variation.

Thus, these analyses do not require separate execution models. They are
different uses of the same fitted AgentSoV dynamics: $\mH_k$ specifies where
abnormal variation can enter, $\mF_k$ determines how it propagates, $\mB_k$
describes how stage inputs influence the system, and $\mC_k$ determines what
parts of that variation become observable.

\section{Additional Experimental Setup}\label{app_exp_set}
\paragraph{Datasets.}
BFCL \citep{liu2026agenthallu,patil2025berkeley} contains 164 tool-calling
trajectories from a single-agent system: 61 are fault-free and 103 contain a
scorable failure. The failures include incorrect tool arguments, missing
required calls, parallel-call conflicts, and unnecessary invocations. We use
five-fold cross-validation over all 164 trajectories.

Magentic-One, drawn from AgentHallu \citep{liu2026agenthallu}, contains 94
trajectories from an orchestrator-with-specialists system comprising an
Orchestrator, WebSurfer, Coder, and ComputerTerminal. Of these, 40 are
fault-free and 54 contain a labeled failure. We use five-fold cross-validation
stratified by fault-free, planning-failure, and execution-failure trajectories.

AEGIS \citep{kong2026aegis} provides official training, validation, and test
splits containing 1749, 454, and 156 MacNet trajectories, respectively. After
removing task overlap across splits and excluding one training trajectory with
a different MacNet configuration, we use 856 training, 364 validation, and 156
test trajectories.
Each trajectory is a topologically ordered DAG with ten calls executed by five
agents and may contain multiple synthetically injected faulty agents.

\paragraph{Baselines.}
Uniform Random ranks the candidate steps or agents in random order. The GPT-5.5
baselines follow the three prompting strategies in
\citet{pmlr-v267-zhang25cq}: All-at-Once supplies the complete trajectory with
attribution instructions, Step-by-Step analyzes each step sequentially, and
Binary Search iteratively narrows the candidate failure window. MASC
\citep{shen2026metacognitive} trains a history-conditioned multilayer
perceptron (MLP) on clean trajectories and uses prediction error as its anomaly
score. OAT \citep{yeh2026tracing} re-encodes trajectory steps with an LLM and
identifies deviations from normal dynamics modeled by a neural controlled
differential equation. We include a baseline only when its released method
provides sufficient implementation detail and supports the corresponding
analysis task.

\paragraph{AgentSoV variants and computation.}
AgentSoV fits its normal dynamics using normal training
executions and performs attribution from the resulting residuals and their
propagation. 
AgentSoV-S uses the same diagnostic representation but, for each
stage, forms features from the standardized local prediction residual and the
change in that residual after adding cross-stage propagation. A class-balanced
logistic classifier shared across stage positions is trained with fault labels
to map these features to the probability that each stage is faulty. 
For specific datasets, we encode text using the all-MiniLM-L6-v2 encoder and reduce the resulting embeddings to 16 dimensions using PCA. The sparsity parameter $\lambda$ in
Eq.~\ref{eq:app_sparse_attribution} is selected independently from $\{0.03,0.1,0.3,1.0,3.0\}$ by maximizing step accuracy.
In virtual-sensor placement, we fix $\eta=10^{-6}$ for numerical
stabilization. 
All
AgentSoV experiments ran on one Apple M3 Pro CPU thread, requiring
no LLM.

\section{Additional Results}\label{app_add_res}

\subsection{Failure Attribution}


\paragraph{The impact of $\mF$.}
Tables~\ref{tab:exp_attribution_shared_f} and
\ref{tab:aegis_attribution_shared_f} examine whether the transition dynamics
can be shared across stages rather than using stage-specific $\mF_k$ across three datasets.
On BFCL, the two parameterizations perform nearly identically: sharing $\mF$
changes AgentSoV-S from $36.9\%$ to $37.7\%$ accuracy and leaves AgentSoV
essentially unchanged at $37.5\%$. This suggests that a common transition
model is sufficient for the relatively homogeneous tool-calling stages.

\begin{table}[htbp]
\caption{Failure attribution on BFCL and Magentic-One using a shared transition matrix $\mF$ across all stages, i.e., $\mF_k = \mF$ for all stage $k$.}
\centering
\resizebox{\linewidth}{!}{
\begin{tabular}{clcccc}
\toprule
\textbf{Dataset} & \textbf{Method} & \textbf{Step Accuracy} & \textbf{Top-3 Step Accuracy} & \textbf{MRR} & \textbf{Agent Accuracy} \\
\midrule

\multirow{2}{*}{\footnotesize BFCL}
& AgentSoV-S 
    & $\mathbf{37.7\%\pm1.3\%}$ 
    &$\mathbf{65.2\%\pm2.7\%}$ & $\mathbf{56.1\%\pm1.4\%}$
    & $\mathbf{37.7\%\pm1.3\%}$ \\
& AgentSoV 
    & $37.5\%\pm1.0\%$  &$63.1\%\pm1.7\%$ & $54.6\%\pm0.5\%$ & $37.5\%\pm1.0\%$ \\
\midrule

\multirow{2}{*}{\footnotesize Magentic-One}
& AgentSoV-S 
    & $32.2\%\pm1.9\%$ 
& $58.5\%\pm1.5\%$ & $46.4\%\pm1.0\%$ & $54.8\%\pm5.1\%$ \\
& AgentSoV
    & $\mathbf{37.8\%\pm2.1\%}$ & $\mathbf{61.5\%\pm1.8\%}$ & $\mathbf{51.0\%\pm1.5\%}$  & $\mathbf{62.6\%\pm0.8\%}$ \\
\bottomrule
\end{tabular}
}
\label{tab:exp_attribution_shared_f}
\end{table}

\begin{table}[htbp]
\centering
\caption{Failure attribution on AEGIS using a shared transition matrix $\mF$ across all stages, i.e., $\mF_k = \mF$ for all stage $k$.}
\label{tab:aegis_attribution_shared_f}
\begin{tabular}{lcccc}
\toprule
\textbf{Method}
& \textbf{Exact-Set Accuracy}
& \textbf{Top-1 Accuracy} 
& \textbf{Top-3 Accuracy} & \textbf{MRR} 
\\
\midrule
AgentSoV-S 
& $38.5\%\pm0.0\%$
& $77.6\%\pm0.3\%$
&$98.2\%\pm0.3\%$ &$87.4\%\pm0.3\%$ 
\\
AgentSoV
& $\mathbf{46.2\%\pm1.3\%}$
& $\mathbf{91.7\%\pm0.5\%}$ & $100.0\%\pm0.0\%$ &$95.4\%\pm0.3\%$ \\
\bottomrule
\end{tabular}
\end{table}

On Magentic-One, AgentSoV remains robust to sharing, with step accuracy changing
from $39.3\%$ to $37.8\%$ and agent accuracy from $61.9\%$ to $62.6\%$.
In contrast, AgentSoV-S decreases from $36.3\%$ to $32.2\%$ in step accuracy
and from $63.0\%$ to $54.8\%$ in agent accuracy, suggesting that its supervised
attribution benefits more from stage-specific transition dynamics.

On AEGIS, sharing $\mF$ improves exact-set accuracy for AgentSoV-S from $33.6\%$ to $38.5\%$, while AgentSoV decreases from
$46.8\%$ to $46.2\%$. AgentSoV improves in top-1 accuracy from $89.7\%$
to $91.7\%$, whereas AgentSoV-S changes from $86.4\%$ to $77.6\%$.
These results indicate that parameter sharing can regularize propagation
estimation when stages exhibit similar transition structure, while
stage-specific $\mF_k$ can be advantageous when propagation patterns are more
heterogeneous.






\paragraph{The impact of $\mH$.}

Table~\ref{tab:exp_attribution_ablation} evaluates the impact of $\mH$ by replacing failure-type-aware disturbance directions ($\mH$) with an identity mapping $\mH{=}\mI$. AgentSoV reduces step accuracy from $37.5\%$ to $35.3\%$
($-2.2\%$), indicating that failure-type-specific disturbance
signatures provide additional discrimination beyond the learned
propagation dynamics.
Together, these ablations show that both components contribute to
failure attribution: $\mF$ provides the cross-stage propagation
structure, while $\mH$ refines how different failure mechanisms
enter the diagnostic state.

\begin{table}[htbp]
\centering
\caption{Failure attribution accuracy on BFCL when replacing the failure-aware disturbance matrix $\mH$ with $\mH{=}\mI$. $\Delta$ denotes the resulting change in accuracy.}
\begin{tabular}{lcc}
\toprule
\textbf{Method} & \textbf{Accuracy} & \textbf{$\Delta$ vs. Failure-Specific $\mH$} \\
\midrule
AgentSoV ($\mF{=}\mathbf{0}$, $\mH{=}\mI$)
    & $26.2\%\pm1.4\%$ & $-1.4\%$  \\
AgentSoV ($\mH{=}\mI$)
    & $35.3\%\pm2.0\%$ & $-2.2\%$ \\
\bottomrule
\end{tabular}
\label{tab:exp_attribution_ablation}
\end{table}



\subsection{Stage-Input Influence}

\begin{figure*}[htbp]
    \centering
    \begin{subfigure}[t]{0.225\textwidth}
        \centering
        \includegraphics[width=\linewidth]{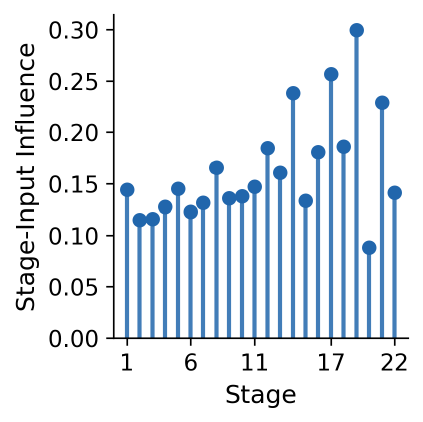}
        \caption{Input influence by stage.}
        \label{fig:bfcl_sensitivity_stage}
    \end{subfigure}\hfill
    \begin{subfigure}[t]{0.225\textwidth}
        \centering
        \includegraphics[width=\linewidth]{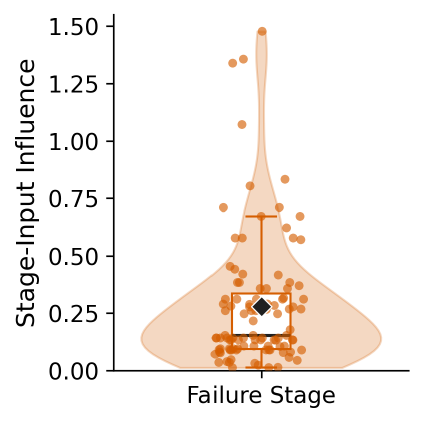}
        \caption{Failure-stage input influence.}
        \label{fig:bfcl_sensitivity_failure}
    \end{subfigure}\hfill
    \begin{subfigure}[t]{0.225\textwidth}
        \centering
        \includegraphics[width=\linewidth]{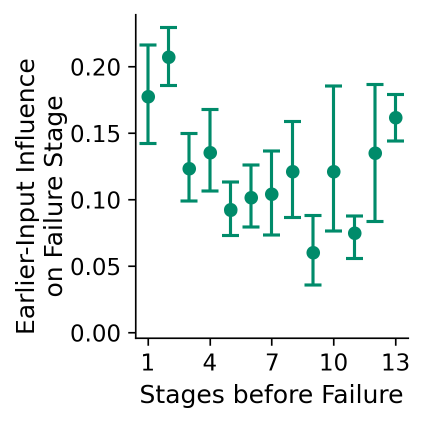}
        \caption{Influence of earlier inputs on failure stage.}
        \label{fig:bfcl_sensitivity_predecessors}
    \end{subfigure}\hfill
    \begin{subfigure}[t]{0.28\textwidth}
        \centering
        \includegraphics[width=\linewidth]{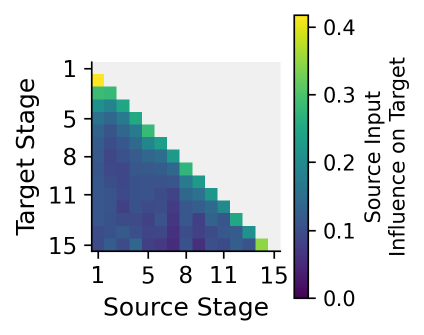}
        \caption{Source-to-target stage influence.}
        \label{fig:bfcl_sensitivity_matrix}
    \end{subfigure}
    \caption{Stage-input influence on BFCL.}
    \label{fig:bfcl_sensitivity_four_panel}
\end{figure*}

Figure~\ref{fig:bfcl_sensitivity_four_panel} shows that stage inputs have
heterogeneous influence on BFCL. In Figure~\ref{fig:bfcl_sensitivity_stage},
the mean influence on the final measurement ranges from approximately $0.11$
to $0.18$ across the well-represented early stages, while several later stages
have larger values. The failure-stage distribution
in Figure~\ref{fig:bfcl_sensitivity_failure} is strongly right-skewed, with a
mean of $0.278$ and a median of $0.153$, showing that a small number of failure
stages exert large downstream effects while many have modest influence.
Figures~\ref{fig:bfcl_sensitivity_predecessors} and
\ref{fig:bfcl_sensitivity_matrix} further show that propagation is largely
local: inputs one and two stages before failure have mean influences of $0.177$
and $0.207$, respectively, whereas most longer-lag means are lower, and the
largest source-to-target effects concentrate near the matrix diagonal. The
decay is not strictly monotonic, indicating that stage separation alone does
not determine influence on BFCL. Overall, AgentSoV reveals which tool-calling inputs
remain consequential downstream and how rapidly their effects are attenuated
across stages.

\end{document}

%% file: math_commands.tex
\usepackage{amsmath,amsfonts,bm}

\def\eqref#1{equation~\ref{#1}}

\def\1{\bm{1}}

\def\va{{\bm{a}}}

\def\ve{{\bm{e}}}

\def\vu{{\bm{u}}}
\def\vv{{\bm{v}}}
\def\vw{{\bm{w}}}
\def\vx{{\bm{x}}}
\def\vy{{\bm{y}}}
\def\vz{{\bm{z}}}

\def\mB{{\bm{B}}}
\def\mC{{\bm{C}}}

\def\mF{{\bm{F}}}
\def\mG{{\bm{G}}}
\def\mH{{\bm{H}}}
\def\mI{{\bm{I}}}

\def\mP{{\bm{P}}}

\def\mSigma{{\bm{\Sigma}}}

\DeclareMathAlphabet{\mathsfit}{\encodingdefault}{\sfdefault}{m}{sl}
\SetMathAlphabet{\mathsfit}{bold}{\encodingdefault}{\sfdefault}{bx}{n}

\newcommand{\R}{\mathbb{R}}

